\documentstyle[myart,12pt]{article}

\normalbaselines
\font\tenbm=cmmib10
\font\sevenbm=cmmib7
\font\fivebm=cmmib5
\newfam\bmfam
\textfont\bmfam=\tenbm \scriptfont\bmfam=\sevenbm
\scriptscriptfont\bmfam=\fivebm
{\count0=\number\bmfam \multiply\count0 by "100
\def\defbgreek#1#2#3{{\count1=\count0 \advance\count1 by "#2#3
  \global\mathchardef#1=\count1 }}
\defbgreek\balpha  0B \defbgreek\brho       1A
\defbgreek\bbeta   0C \defbgreek\bsigma     1B
\defbgreek\bgamma  0D \defbgreek\btau       1C
\defbgreek\bdelta  0E \defbgreek\bupsilon   1D
\defbgreek\bepsilon0F \defbgreek\bphi       1E
\defbgreek\bzeta   10 \defbgreek\bchi       1F
\defbgreek\bmeta   11 \defbgreek\bpsi       20
\defbgreek\btheta  12 \defbgreek\bomega     21
\defbgreek\biota   13 \defbgreek\bvarepsilon22
\defbgreek\bkappa  14 \defbgreek\bvartheta  23
\defbgreek\blambda 15 \defbgreek\bvarpi     24
\defbgreek\bmu     16 \defbgreek\bvarrho    25
\defbgreek\bnu     17 \defbgreek\bvarsigma  26
        \defbgreek\bxi     18 \defbgreek\bvarphi    27
\defbgreek\bpi     19}
\input tcilatex.tex

\begin{document}

\author{Yuri A. Rylov}
\title{Anderson's absolute objects and constant timelike vector hidden in Dirac
matrices.}
\date{Institute for Problems in Mechanics, Russian Academy of Sciences \\
101-1 ,Vernadskii Ave., Moscow, 117526, Russia \\
email: rylov@ipmnet.ru\\
Web site: {$http://rsfq1.physics.sunysb.edu/\symbol{126}rylov/yrylov.htm$}\\
or mirror Web site: {$http://194.190.131.172/\symbol{126}rylov/yrylov.htm$}}
\maketitle

\begin{abstract}
Anderson's theorem asserting, that symmetry of dynamic equations written in
the relativisitically covariant form is determined by symmetry of its
absolute objects, is applied to the free Dirac equation. $\gamma $-matrices
are the only absolute objects of the Dirac equation. There are two ways of
the $\gamma $-matrices transformation: (1) $\gamma ^{i}$ is a 4-vector and $%
\psi $ is a scalar, (2) $\gamma ^{i}$ are scalars and $\psi $ is a spinor.
In the first case the Dirac equation is nonrelativistic, in the second one
it is relativistic. Transforming Dirac equation to another scalar--vector
variables, one shows that the first way of transformation is valid, and the
Dirac equation is not relativistic completely. \ 
\end{abstract}



\section{Introduction}

Relativistic covariance of dynamic equations and its role in relativistic
physics was discussed intensively in the sixth decade of XX century. It
seems now that all problems of relativistic description of relativistic
dynamical systems have been discussed and solved. Unfortunately, it is not
so. Some problems remain. In particular, there is a problem, connected with
application of so called absolute objects.

Concept of the absolute object was introduced, apparently, by J.L. Anderson 
\cite{A67}, who divided all objects, connected with dynamic systems, into
two sorts: dynamical objects and absolute objects.

Dynamical objects (variables) are such objects, which are different for
different solutions of dynamic equations. The absolute object is such an
object, which is the same for all solutions of the dynamic equations \cite
{A67}.

For instance, let us consider a system of Maxwell equations, describing
electromagnetic field tensor $F^{ik}$, generated by a given 4-current $J^{i}$%
.

\begin{equation}
\partial _{k}F^{ik}=4\pi J^{i},\qquad \varepsilon _{iklm}g^{jm}\partial
_{j}F^{kl}=0,\qquad \partial _{k}\equiv \frac{\partial }{\partial x^{k}}
\label{b1.1}
\end{equation}
The 4-current is considered to be a given function $J^{i}=J^{i}\left(
x\right) $ of coordinates $x$ of inertial coordinate system $K$. Here the
electromagnetic field tensor $F^{ik}$ is a dynamical object. The
Levi-Chivita pseudotensor $\varepsilon _{iklm},$ the metric tensor $g^{jm}$
and external 4-current $J^{i}\left( x\right) $ are absolute objects, because
they are the same for all solutions of dynamic equations (\ref{b1.1}).

If one considers the metric tensor $g^{ik}$ to be a solution of the
gravitation equation (but not as a fixed quantity), the metric tensor stops
to be an absolute object and becomes to be a dynamical object. Similarly, if
the 4-current $J^{i}$ is determined by charged particles, whose motion is
described by some dynamic equations, $J^{i}$ becomes to be a dynamical
object.

It is a common practice to think that if dynamic equations of a system can
be written in the relativistically covariant form, such a possibility
provides automatically a relativistic character of considered dynamic
system, described by these equations. In general, it is valid only in the
case, when dynamic equations do not contain absolute objects, or these
absolute objects has the Poincare group as a group of their symmetry \cite
{A67}. More exactly, J.L. Anderson shows that the symmetry group of a system
of dynamic equations, written in the relativistically covariant form,
coincides with the group of symmetry of all absolute objects of this system.

The absolute object, which is a constant unit timelike vector $%
l_{k},\;\;k=0,1,2,3$ 
\begin{equation}
g^{ik}l_{i}l_{k}=1,\qquad l_{k}=\text{const.}  \label{a6.3}
\end{equation}
is of most interest. Such a vector $l_{k}$ is to be interpreted, as a
vector, describing a preferred direction in the space-time. Existence of a
preferred direction in the space-time is incompatible with the relativity
principles.

Let us consider equations 
\begin{equation}
m\frac{d^{2}x^{\alpha }}{dt^{2}}=eF_{.0}^{\alpha }+eF_{.\beta }^{\alpha }%
\frac{dx}{dt}^{\beta },\qquad \alpha =1,2,3;  \label{a6.11}
\end{equation}
\[
\frac{m}{2}\frac{d}{dt}(\frac{dx}{dt}^{\alpha }\frac{dx}{dt}^{\alpha
})=eF_{.0}^{\alpha }\frac{dx}{dt}^{\alpha }, 
\]
describing motion of a nonrelativistic particle of the mass $m$ and of the
charge $e$ in the given electromagnetic field $F^{ik}$. The speed of the
light is chosen $c=1$ Equations (\ref{a6.11}) are written in the
non-covariant form, and they are incompatible with the relativity
principles. Introducing a constant unit vector $l_{k}$, one can write four
equations (\ref{a6.11}) in the relativistically covariant form 
\begin{equation}
m\frac{d}{d\tau }\left[ \frac{\dot{x}^{i}}{l_{k}\dot{x}^{k}}-{\frac{1}{2}}%
g^{ik}l_{k}\frac{\dot{x}^{s}g_{sl}\dot{x}^{l}}{(l_{j}\dot{x}^{j})^{2}}\right]
=eF^{il}g_{lk}\dot{x}^{k};\qquad i=0,1,2,3  \label{a6.2}
\end{equation}
\[
\dot{x}^{k}\equiv \frac{dx^{k}}{d\tau },\qquad g_{ik}=\text{diag}\left\{
1,-1,-1,-1\right\} 
\]
where $\tau $ is a parameter along the world line $x^{l}=x^{l}\left( \tau
\right) ,\;\;l=0,1,2,3$ of the particle, and $F^{il}$ is some fixed function
of coordinates $x$.

The equation (\ref{a6.2}) is relativistically covariant with respect to
vectors $x^{i},$ $l_{i},$ and tensors $F^{ik},$ $g_{ik}.$ A reference to the
quantities $x^{i},$ $l_{i},$ $F^{ik},$ $g_{ik}$ means that they are
considered to be formal variables (but not functions of coordinates $x$).

Transforming quantities $x^{i},$ $l_{i},$ $F^{ik},$ $g_{ik}$ from the
coordinate $K$ to the coordinate system \ $\tilde{K}$ 
\begin{equation}
x^{i}\rightarrow \tilde{x}^{i}=x^{i}+\omega _{.k}^{i}x^{k}+o\left( \omega
\right) ,\qquad l_{i}\rightarrow \tilde{l}_{i}=\frac{\partial x^{k}}{%
\partial \tilde{x}^{i}}l_{k}  \label{b1.2}
\end{equation}
\begin{equation}
F^{ik}\rightarrow \tilde{F}^{ik}=\frac{\partial \tilde{x}^{i}}{\partial x^{l}%
}\frac{\partial \tilde{x}^{k}}{\partial x^{m}}F^{lm}\qquad g^{ik}\rightarrow 
\tilde{g}^{ik}=\frac{\partial \tilde{x}^{i}}{\partial x^{l}}\frac{\partial 
\tilde{x}^{k}}{\partial x^{m}}g^{lm}=g^{ik},  \label{b1.3}
\end{equation}
one obtains instead of (\ref{a6.2}) 
\begin{equation}
m\frac{d}{d\tau }\left[ \left( \tilde{l}_{k}\frac{d\tilde{x}^{k}}{d\tau }%
\right) ^{-1}\frac{d\tilde{x}^{i}}{d\tau }-{\frac{1}{2}}g^{ik}\tilde{l}%
_{k}\left( \tilde{l}_{j}\frac{d\tilde{x}^{j}}{d\tau }\right) ^{-2}g_{sl}%
\frac{d\tilde{x}^{s}}{d\tau }\frac{d\tilde{x}^{l}}{d\tau }\right] =e\tilde{F}%
^{il}g_{lk}\frac{d\tilde{x}^{k}}{d\tau }  \label{c1.4}
\end{equation}
Equations (\ref{a6.2}) and (\ref{c1.4}) have the same form, provided the
quantities $x^{i},$ $l_{i},$ $F^{ik}$ are considered to be formal variables.
If for instance, the electromagnetic field $F^{ik}$ is considered to be a
function of coordinates $x$, i.e. $F^{ik}=F^{ik}\left( x\right) $, then $%
F^{ik}\left( x\right) $ and $\tilde{F}^{ik}\left( \tilde{x}\right) $ are
different functions respectively of $x$ and $\tilde{x}$. In this case the
equations (\ref{a6.2}) and (\ref{c1.4}) have different form, because rhs of (%
\ref{a6.2}) and (\ref{c1.4}) are different function of $x$ and $\tilde{x}$
respectively. In this case one must say that the equation (\ref{a6.2}) is
not relativistically covariant with respect to the quantities $x^{i},$ $%
l_{i},$ (now a reference to the variable $F^{ik}$ is absent, and it is
considered to be a function of $x$).

Thus, the equation (\ref{a6.2}) is relativistically covariant with respect
to the quantities $x^{i},$ $l_{i},$ $F^{ik}.$ Nevertheless it is
incompatible with the relativity principles. Now the reason of this
incompatibility is an existence of the constant timelike unit vector $l_{k}.$
This vector describes a preferred direction in space-time. Any 3-plane
orthogonal to $l_{k}$ may be considered as set of simultaneous events. If
the coordinate system is chosen in such a way, that the vector $l_{k}$ takes
the form $l_{k}=\left\{ 1,0,0,0\right\} $,\ \ $t=x^{0}=\tau $, the equation (%
\ref{a6.2}) takes the form (\ref{a6.11}).

Thus, the nonrelativistic character of the equation may be described either
by non-covariant form of the equation, or by introducing the absolute object 
$l_{k},$ whose symmetry group is a subgroup of the Lorentz group and does
not coincide with the Lorentz group. If a system of dynamic equation is
written in a relativistically covariant form and contains a constant
timelike unit vector $l_{k}$. This vector describes a split of the
space-time into space and time, and the system of dynamic equations is
incompatible with the relativity principles.

\section{Free Dirac equation}

Let ${\cal S}_{{\rm D}}$ be the dynamic system, described by the free Dirac
equation 
\begin{equation}
i\hbar \gamma ^{l}\partial _{l}\psi -m\psi =0  \label{f1.1}
\end{equation}
which can be obtained from the action 
\begin{equation}
{\cal S}_{{\rm D}}:\qquad {\cal A}_{{\rm D}}[\bar{\psi},\psi ]=\int (-m\bar{%
\psi}\psi +{\frac{i}{2}}\hbar \bar{\psi}\gamma ^{l}\partial _{l}\psi -{\frac{%
i}{2}}\hbar \partial _{l}\bar{\psi}\gamma ^{l}\psi )d^{4}x  \label{f1.0}
\end{equation}
Here $\psi $ is four-component complex wave function, $\bar{\psi}=\psi
^{\ast }\gamma ^{0}$ is conjugate wave function, and $\psi ^{\ast }$ is the
Hermitian conjugate one. $\gamma ^{i}$, $i=0,1,2,3$ are $4\times 4$ complex
constant matrices, satisfying the relations 
\begin{equation}
\gamma ^{l}\gamma ^{k}+\gamma ^{k}\gamma ^{l}=2g^{kl}I,\qquad k,l=0,1,2,3.
\end{equation}
where $I$ is unit $4\times 4$ matrix. The speed of the light is chosen $c=1$%
.The quantities $\gamma ^{l}$ form an absolute object, because they are
similaar for all solutions of the Dirac equation (\ref{f1.1})

There are two approaches to the Dirac equation. In the first approach \cite
{S30,S51} the wave function $\psi $ is considered to be a scalar function
defined on the field of Clifford numbers $\gamma ^{l}$, 
\begin{equation}
\psi =\psi (x,\gamma )\Gamma ,\qquad \overline{\psi }=\Gamma \overline{\psi }%
(x,\gamma ),  \label{b1.4}
\end{equation}
where $\Gamma $ is a constant nilpotent factor which has the property $%
\Gamma f(\gamma )\Gamma =a\Gamma $. Here $f(\gamma )$ is arbitrary function
of $\gamma ^{l}$ and $a$ is a complex number, depending on the form of the
function $f$. Within such an approach $\psi $, $\bar{\psi}$ transform as
scalars and $\gamma ^{l}$ transform as components of a 4-vector under the
Lorentz transformations. In this case the symmetry group of $\gamma ^{l}$ is
a subgroup of the Lorentz group, and ${\cal S}_{{\rm D}}$ is nonrelativistic
dynamic system. Then the matrix vector $\gamma ^{l}$ describes some
preferred direction in the space-time.

In the second (conventional) approach \cite{S61} $\psi $ is considered to be
a spinor, and $\gamma ^{l},\quad l=0,1,2,3$ are scalars with respect to the
transformations of the Lorentz group. In this case the symmetry group of the
absolute objects $\gamma ^{l}$ is the Lorentz group, and dynamic system $%
{\cal S}_{{\rm D}}$ is considered to be a relativistic dynamic system.

Of course, the approaches leading to incompatible conclusions cannot be both
valid. At least, one of them is wrong. Analyzing the two approaches,
Sommerfeld \cite{S51} considered the first approach to be more reasonable.
In the second case the analysis is rather difficult due to non-standard
transformations of $\gamma ^{l}$ and $\psi $ under linear coordinate
transformations $T$. Indeed, the transformation $T$ for the vector $j^{l}=%
\bar{\psi}\gamma ^{l}\psi $ has the form 
\begin{equation}
\tilde{\overline{\psi }}\tilde{\gamma}^{l}\tilde{\psi}=\frac{\partial \tilde{%
x}^{l}}{\partial x^{s}}\bar{\psi}\gamma ^{s}\psi ,  \label{b1.5}
\end{equation}
where quantities marked by tilde mean quantities at the transformed
coordinate system. This transformation can be carried out by two different
ways 
\begin{equation}
1:\;\;\;\tilde{\psi}=\psi ,\qquad \tilde{\overline{\psi }}=\overline{\psi }%
,\qquad \tilde{\gamma}^{l}=\frac{\partial \tilde{x}^{l}}{\partial x^{s}}%
\gamma ^{s},\qquad l=0,1,2,3  \label{b1.6}
\end{equation}
\begin{equation}
2:\;\;\;\tilde{\gamma}^{l}=\gamma ^{l},\qquad l=0,1,2,3,\qquad \tilde{\psi}%
=S(\gamma ,T)\psi ,\qquad \tilde{\overline{\psi }}=\overline{\psi }%
S^{-1}(\gamma ,T),  \label{b1.7}
\end{equation}
\begin{equation}
S^{\ast }(\gamma ,T)\gamma ^{0}=\gamma ^{0}S^{-1}(\gamma ,T)  \label{b1.8}
\end{equation}
The relations (\ref{b1.6}) correspond to the first approach and the
relations (\ref{b1.7}) correspond to the second one. Both ways (\ref{b1.6})
and (\ref{b1.7}) lead to the same result, provided 
\begin{equation}
S^{-1}(\gamma ,T)\gamma ^{l}S(\gamma ,T)=\frac{\partial \tilde{x}^{l}}{%
\partial x^{s}}\gamma ^{s}  \label{b1.9}
\end{equation}
In particular, for infinitesimal Lorentz transformation 
\[
x^{i}\rightarrow x^{i}+\delta \omega _{.k}^{i}x^{k} 
\]
$S(\gamma ,T)$ has the form \cite{S61} 
\begin{equation}
S(\gamma ,T)=\exp \left( \frac{\delta \omega _{ik}}{8}\left( \gamma
^{i}\gamma ^{k}-\gamma ^{k}\gamma ^{i}\right) \right)  \label{b1.10}
\end{equation}
The second way (\ref{b1.7}) has a defect. The transformation law of $\psi $
depends on $\gamma $, i.e. under linear coordinate transformation $T$ the
components of $\psi $ transform through $\psi $ and $\gamma ^{l}$, but not
only through $\psi $. Note that tensor components at a coordinate system
transform only through tensor components at other coordinate system, and
this transformation does not contain any absolute objects. (for instance,
the relation (\ref{b1.5})).

The fact that the symmetry group of a dynamic system coincides with the
symmetry group of absolute objects was derived at the supposition, that
under the coordinate transformation any object transforms only via its
components. This condition is violated in the second case, and one cannot be
sure that the symmetry group of dynamic system coincides with that of
absolute objects.

\section{The case of two-dimensional space-time}

To determine which of the two approaches is valid, let us consider such a
transformation of the dependent variable $\psi ,$ which eliminates the $%
\gamma $-matrices. At first, we consider a more simple case of the
two-dimensional space-time. Let 
\begin{equation}
\psi _{{\rm D}}=(_{\psi _{-}}^{\psi _{+}}),\qquad \bar{\psi}_{{\rm D}}=\psi
_{{\rm D}}^{\ast }\gamma ^{0},\qquad \psi _{{\rm D}}^{\ast }=(\psi
_{+}^{\ast },\psi _{-}^{\ast })  \label{m1.1}
\end{equation}
\begin{equation}
\gamma ^{0}=\left( 
\begin{array}{cc}
0 & 1 \\ 
1 & 0
\end{array}
\right) ,\qquad \gamma ^{1}=\left( 
\begin{array}{cc}
0 & 1 \\ 
-1 & 0
\end{array}
\right) .  \label{m1.2}
\end{equation}

Representation (\ref{m1.2}) of $\gamma $-matrices is chosen in such a way
that the pseudo-scalar matrix $\gamma ^{0}\gamma ^{1}$ be diagonal, and the
wave functions $\psi _{{\rm D}}=(_{0}^{\psi _{+}}),\quad \psi _{{\rm D}%
}=(_{\psi _{-}}^{0})$ be its eigenfunctions for any choice of $\psi
_{+},\psi _{-}$. In this case in virtue of (\ref{m1.1}) the Dirac equation 
\begin{equation}
i\hbar \gamma ^{l}\partial _{l}\psi _{{\rm D}}-m\psi _{{\rm D}}=0
\label{m1.3}
\end{equation}
takes the form 
\begin{equation}
\psi _{+}=i\lambda \partial _{+}\psi _{-},\qquad \psi _{-}=i\lambda \partial
_{-}\psi _{+},  \label{m1.4}
\end{equation}
\begin{equation}
\lambda \equiv \hbar /m,\qquad \partial _{\pm }\equiv \partial _{0}\pm
\partial _{1}  \label{m1.5}
\end{equation}
It follows from Eq.(\ref{m1.4}) that both wave functions $\psi _{\pm }$
satisfy the free Klein-Gordon equation 
\begin{equation}
\lambda ^{2}\partial _{l}\partial ^{l}\psi _{\pm }+\psi _{\pm }=0
\label{m1.6}
\end{equation}

Let us introduce the two-component differential operator 
\begin{equation}
\hat{{\cal L}}(w,\partial ,\lambda )=\left( 
\begin{array}{c}
\sqrt{w_{+}}+i\lambda \sqrt{w_{-}}\partial _{+} \\ 
\sqrt{w_{-}}+i\lambda \sqrt{w_{+}}\partial _{-}
\end{array}
\right) .  \label{m1.7}
\end{equation}
where $w_{l}=(w_{0},w_{1})$ is a constant timelike vector and 
\[
w_{+}=w_{0}+w_{1},\qquad w_{-}=w_{0}-w_{1} 
\]
Under the continuous Lorentz transformation 
\begin{equation}
\begin{array}{c}
x^{0}\rightarrow \tilde{x}^{0}=x^{0}\cosh \chi +x^{1}\sinh \chi \\ 
x^{1}\rightarrow \tilde{x}^{1}=x^{1}\cosh \chi +x^{0}\sinh \chi
\end{array}
\label{m1.8}
\end{equation}
the components $w_{\pm }$ and $\partial _{\pm }$ transform as follows 
\begin{equation}
\begin{array}{c}
w_{+}\rightarrow \tilde{w}_{+}=e^{\chi }w_{+},\qquad w_{-}\rightarrow \tilde{%
w}_{-}=e^{-\chi }w_{-} \\ 
\partial _{+}\rightarrow \tilde{\partial}_{+}=e^{\chi }\partial _{+},\qquad
\partial _{-}\rightarrow \tilde{\partial}_{-}=e^{-\chi }\partial _{-}
\end{array}
\label{m1.9}
\end{equation}
According to Eqs. (\ref{m1.8}), (\ref{m1.9}) the differential operator (\ref
{m1.7}) transforms as follows 
\[
\hat{{\cal L}}(w,\partial ,\lambda )\rightarrow \hat{{\cal L}}(\tilde{w},%
\tilde{\partial},\lambda )=\left( 
\begin{array}{c}
\sqrt{\tilde{w}_{+}}+i\lambda \sqrt{\tilde{w}_{-}}\tilde{\partial}_{+} \\ 
\sqrt{\tilde{w}_{-}}+i\lambda \sqrt{\tilde{w}_{+}}\tilde{\partial}_{-}
\end{array}
\right) = 
\]
\begin{equation}
=\left( 
\begin{array}{c}
e^{\chi /2}(\sqrt{w_{+}}+i\lambda \sqrt{w_{-}}\partial _{+}) \\ 
e^{-\chi /2}(\sqrt{w_{-}}+i\lambda \sqrt{w_{+}}\partial _{-})
\end{array}
\right) =e^{-\gamma ^{0}\gamma ^{1}\chi /2}\hat{{\cal L}}(w,\partial
,\lambda )  \label{m1.10}
\end{equation}
Under the space reflection 
\begin{equation}
x^{0}\rightarrow \tilde{x}^{0}=x^{0},\qquad x^{1}\rightarrow \tilde{x}%
^{1}=-x^{1}  \label{m1.11}
\end{equation}
one has 
\begin{equation}
\begin{array}{c}
w_{+}\rightarrow \tilde{w}_{+}=w_{-},\qquad w_{-}\rightarrow \tilde{w}%
_{-}=w_{+} \\ 
\partial _{+}\rightarrow \tilde{\partial}_{+}=\partial _{-},\qquad \partial
_{-}\rightarrow \tilde{\partial}_{-}=\partial _{+}
\end{array}
\label{m1.12}
\end{equation}
\begin{equation}
\hat{{\cal L}}(w,\partial ,\lambda )\rightarrow \hat{{\cal L}}(\tilde{w},%
\tilde{\partial},\lambda )=c\gamma ^{0}\hat{{\cal L}}(w,\partial ,\lambda )
\label{m1.13}
\end{equation}

Under the time reflection 
\begin{equation}
x^{0}\rightarrow \tilde{x}^{0}=-x^{0},\qquad x^{1}\rightarrow \tilde{x}%
^{1}=x^{1}  \label{m1.14}
\end{equation}
one can write 
\begin{equation}
\begin{array}{c}
w_{+}\rightarrow \tilde{w}_{+}=e^{i\pi }w_{-},\qquad w_{-}\rightarrow \tilde{%
w}_{-}=e^{-i\pi }w_{+} \\ 
\partial _{+}\rightarrow \tilde{\partial}_{+}=e^{i\pi }\partial _{-},\qquad
\partial _{-}\rightarrow \tilde{\partial}_{-}=e^{-i\pi }\partial _{+}
\end{array}
\label{m1.15}
\end{equation}
\begin{equation}
\hat{{\cal L}}(w,\partial ,\lambda )\rightarrow \hat{{\cal L}}(\tilde{w},%
\tilde{\partial},\lambda )=e^{i\pi /2}\gamma ^{1}\hat{{\cal {L}}}(w,\partial
,\lambda )  \label{m1.16}
\end{equation}
It means that the differential operator $\hat{{\cal {L}}}(w,\partial
,\lambda )$ transforms as a spinor under all transformations of the Lorentz
group.

Let us form the two-component quantity 
\begin{equation}
\psi _{{\rm D}}=(_{\psi _{-}}^{\psi _{+}})=\hat{{\cal L}}(w,\partial
,\lambda )\psi  \label{m1.17}
\end{equation}
If $\psi $ is a scalar, satisfying the Klein-Gordon equation 
\begin{equation}
\lambda ^{2}\partial _{l}\partial ^{l}\psi +\psi =0,  \label{m1.18}
\end{equation}
then $\psi _{{\rm D}}$ is a spinor, satisfying the Dirac equation (\ref{m1.3}%
) for any choice of the timelike constant vector $w_{l}=(w_{0},w_{1})$. Vice
versa, if the spinor $\psi _{{\rm D}}$ satisfies the Dirac equation (\ref
{m1.3}), then the scalar $\psi $ defined by Eq.(\ref{m1.17}) satisfies the
Klein-Gordon equation (\ref{m1.18}) for any choice of the timelike vector $w$%
.

Let us compare equations (\ref{m1.3}) and (\ref{m1.18}). None of them
contains the vector $w$ explicitly, but connection (\ref{m1.17}) between $%
\psi $ and $\psi _{{\rm D}}$ contains this vector $w.$ This fact can be
explained only by the fact that the vector $w$ is ''hidden'' inside the $%
\gamma $-matrices. Eliminating $\gamma $-matrices by means of a changing of
variables in (\ref{m1.3}), one discovers the constant timelike vector. Let
us show this.

The action for the dynamic system ${\cal S}_{{\rm D}}$ has the form 
\begin{equation}
{\cal S}_{{\rm D}}:\qquad {\cal A}_{{\rm D}}[\bar{\psi},\psi ]=\int (-m\bar{%
\psi}\psi +{\frac{i}{2}}\hbar \bar{\psi}\gamma ^{l}\partial _{l}\psi -{\frac{%
i}{2}}\hbar \partial _{l}\bar{\psi}\gamma ^{l}\psi )d^{2}x  \label{m1.19}
\end{equation}
Let us substitute four real components of the two-component complex wave
function $\psi $ by four scalar-vector variables $\rho $, $j^{i},$ $\varphi $

\begin{equation}
\rho =\bar{\psi}\psi ,\qquad j^{i}=\bar{\psi}\gamma ^{i}\psi ,\qquad i=0,1
\label{m1.20}
\end{equation}
Let us set 
\begin{equation}
\gamma ^{0}=\sigma _{1},\;\;\;\gamma ^{1}=i\sigma _{2}  \label{m1.21}
\end{equation}
where ${\bf \sigma }=\left\{ \sigma _{1},\sigma _{2},\sigma _{3}\right\} $
are Pauli matrices, having the property 
\begin{equation}
\sigma _{\alpha }\sigma _{\beta }=\sigma _{0}\delta _{\alpha \beta
}+i\varepsilon _{\alpha \beta \gamma }\sigma _{\gamma },\qquad \alpha ,\beta
=1,2,3  \label{m1.22}
\end{equation}
Here $\varepsilon _{\alpha \beta \gamma }$ is the Levi-Chivita
3-pseudotensor $\varepsilon _{123}=1$, $\sigma _{0}$ is the unite matrix .
Let us represent $\psi $ in the form 
\begin{equation}
\psi =A\left( {\bf \sigma n}\right) e^{i\varphi }\Pi ,\qquad \bar{\psi}=A\Pi
\left( {\bf \sigma n}\right) \sigma _{1}e^{-i\varphi },\qquad \left( {\bf %
\sigma n}\right) \equiv \sigma _{\alpha }n_{\alpha },\qquad {\bf n}%
^{2}=n_{\alpha }n_{\alpha }=1  \label{m1.23}
\end{equation}
\begin{equation}
\Pi ={\frac{1}{2}}(1+\gamma ^{0})={\frac{1}{2}}(1+\sigma _{1}),
\label{m1.24}
\end{equation}
where $A$, $\;{\bf n}=\left\{ n_{1},n_{2},n_{3}\right\} $ are intermediate
variables, which will be expressed via variables $\rho $, $j^{i}$. $\Pi $ is
the zero divisor. Using identity (\ref{m1.22}) and its corollary 
\begin{equation}
\left( {\bf \sigma n}\right) \sigma _{\alpha }\left( {\bf \sigma n}\right)
\equiv -{\bf n}^{2}\sigma _{\alpha }+2n_{\alpha }\left( {\bf \sigma n}\right)
\label{m1.25}
\end{equation}
one obtains 
\begin{equation}
\rho =\bar{\psi}\psi =A^{2}\left( -1+2n_{1}^{2}\right) \qquad
j^{0}=A^{2}\qquad j^{1}=-2A^{2}n{\bf _{3}}n{\bf _{1}}  \label{m1.26}
\end{equation}

Resolving (\ref{m1.26}) with respect to components of the 3-vector ${\bf n}$
and taking into account that ${\bf n}^{2}=1$, one obtains 
\begin{equation}
n_{1}=\sqrt{\frac{j^{0}+\rho }{2j^{0}}},\qquad n_{2}=\sqrt{\frac{%
j^{i}j_{i}-\rho ^{2}}{2j^{0}\left( j^{0}+\rho \right) }},\qquad n_{3}=-\frac{%
j^{1}}{\sqrt{2j^{0}\left( j^{0}+\rho \right) }}.  \label{m1.27}
\end{equation}

Let us calculate Lagrangian density ${\cal L}$ of the action (\ref{m1.19})
in terms of components of the vector ${\bf n}$. One obtains 
\begin{equation}
{\cal L}=-m\rho -\hbar A^{2}n_{2}^{2}\left( \partial _{0}\frac{n_{3}}{n_{2}}%
-\partial _{1}\frac{n_{1}}{n_{2}}\right) -\hbar j^{i}\partial _{i}\varphi
\label{m1.29}
\end{equation}

Substituting $A^{2}=j^{0}$ and (\ref{m1.27}) into (\ref{m1.29}), one derives 
\begin{equation}
{\cal L}=-m\rho +\hbar \frac{j^{i}j_{i}-\rho ^{2}}{\left( j^{0}+\rho \right) 
}\left( \partial _{0}\frac{j^{1}}{\sqrt{\left( j^{i}j_{i}-\rho ^{2}\right) }}%
+\partial _{1}\frac{j^{0}+\rho }{\sqrt{j^{i}j_{i}-\rho ^{2}}}\right) -\hbar
j^{i}\partial _{i}\varphi  \label{m1.30}
\end{equation}
The Lagrangian density is expressed in terms of two scalars $\rho $,$\varphi 
$ and the vector $j^{i}$. ${\cal L}$ is written in non-covariant form. It is
not clear, if it possible to transform it to relativistically covariant
form. To show that it is possible, let us introduce the two-component
quantities 
\begin{equation}
q^{l}=\frac{j^{l}+\rho f^{l}}{\sqrt{j^{i}j_{i}-\rho ^{2}}},\qquad
l=0,1,\qquad f^{l}=\left\{ 1,0\right\}  \label{m1.31}
\end{equation}

Now resolving relations (\ref{m1.31}) with respect to $j^{i}$ in the form 
\begin{equation}
j^{l}=\frac{2\rho \left( q^{s}f_{s}\right) }{\left( q^{k}q_{k}-1\right) }%
q^{l}-\rho f^{l},\;\;\;\;\;l=0,1  \label{m1.32}
\end{equation}
and substituting $j^{i}$ in (\ref{m1.30}), one obtains expression for the
action 
\begin{equation}
{\cal S}_{{\rm D}}:\qquad {\cal A}_{{\rm D}}[\rho ,\varphi ,j]=\int (-m\rho
-\hbar \frac{2\rho \left( \partial _{0}q_{1}-\partial _{1}q_{0}\right) }{%
\left( q^{s}q_{s}-1\right) }-\hbar j^{i}\partial _{i}\varphi )d^{2}x
\label{m1.33}
\end{equation}
where $q^{l}$ is expressed via dependent dynamical variables $\rho ,\varphi
,j^{i}$ by means of the relation (\ref{m1.31}). If $\ f^{i}$ is  a vector,
then accorfng to (\ref{m1.31}) $q^{i}$ is also vector and the Lagrangian
density in (\ref{m1.33}) has the covariant form, ${\cal L}$ is an invariant.

Thus, eliminating $\gamma $-matrices, and writing the Lagrangian density in
the relativistically covariant form, one discovers the constant timelike
unite vector $f^{i}$. This vector is an absolute object, describing a
preferred space-time direction, that is incompatible with the relativity
principles.

\section{The case of four-dimensional space-time}

A similar elimination of Dirac matrices can be made in the case of the
four-dimensi\-onal space-time. The state of dynamic system ${\cal S}_{{\rm D}%
}$ (\ref{f1.0}) is described by eight real dependent variables (eight real
components of four-component complex wave function $\psi $). It is possible
to transform the variables $\psi $ and to describe this system in terms of
scalar-vector variables $j^{l},S^{l}$, $(l=0,1,2,3)$, $\varphi ,\kappa $.
The current 4-vector $j^{l}$ and the spin 4-pseudovector $S^{l}$ are defined
by the relations

\[
j^{l}=\bar{\psi}\gamma ^{l}\psi ,\qquad l=0,1,2,3,\qquad \bar{\psi}=\psi
^{\ast }\gamma ^{0}; 
\]
\begin{equation}
S^{l}=i\bar{\psi}\gamma _{5}\gamma ^{l}\psi ,\qquad l=0,1,2,3,\qquad \gamma
_{5}=\gamma ^{0123}\equiv \gamma ^{0}\gamma ^{1}\gamma ^{2}\gamma ^{3};
\label{f1.13a}
\end{equation}
The scalar $\varphi $ and pseudoscalar $\kappa $ are defined implicitly via
the wave function $\psi $ by the relations 
\begin{equation}
\psi =Ae^{i\varphi +{\frac{1}{2}}\gamma _{5}\kappa }e^{-{\frac{i}{2}}\gamma
_{5}\bsigma\bmeta}e^{{\frac{i\pi }{2}}\bsigma {\bf n}}\Pi  \label{f1.11}
\end{equation}
\begin{equation}
\psi ^{\ast }=A\Pi e^{-{\frac{i\pi }{2}}\bsigma {\bf n}}e^{-{\frac{i}{2}}%
\gamma _{5}\bsigma\bmeta}e^{-i\varphi -{\frac{1}{2}}\gamma _{5}\kappa }
\label{f1.12}
\end{equation}
where (*) means the Hermitian conjugation.

One uses the following designations 
\begin{equation}
{\bf \sigma }=\{\sigma _{1},\sigma _{2},\sigma _{3},\}=\{-i\gamma ^{2}\gamma
^{3},-i\gamma ^{3}\gamma ^{1},-i\gamma ^{1}\gamma ^{2}\}  \label{f1.10}
\end{equation}
\begin{equation}
\Pi ={\frac{1}{4}}(1+\gamma ^{0})(1+{\bf z\sigma }),\qquad {\bf z}%
=\{z^{\alpha }\}=\hbox{const},\qquad \alpha =1,2,3;\qquad {\bf z}^{2}=1
\label{f1.13}
\end{equation}
The quantities $A$, $\kappa $, $\varphi $, ${\bf \eta }=\{\eta ^{\alpha }\}$%
, ${\bf n}=\{n^{\alpha }\}$, $\alpha =1,2,3,\;$ ${\bf n}^{2}=1$ are eight
real parameters, determining the wave function $\psi .$ These parameters may
be considered as new dependent variables, describing the state of dynamic
system ${\cal S}_{{\rm D}}$. The quantity $\varphi $ is a scalar, and $%
\kappa $ is a pseudoscalar.

Six remaining variables $A,$ ${\bf \eta }=\{\eta ^{\alpha }\}$, ${\bf n}%
=\{n^{\alpha }\}$, $\alpha =1,2,3,\;$ ${\bf n}^{2}=1$ are intermediate. They
can be expressed through the current 4-vector $j^{l}=\bar{\psi}\gamma
^{l}\psi $ and spin 4-pseudovector $S^{i}$, defined by the relation (\ref
{f1.13a}). Because of two identities 
\begin{equation}
S^{l}S_{l}\equiv -j^{l}j_{l},\qquad j^{l}S_{l}\equiv 0.  \label{f1.14}
\end{equation}
there is only six independent components among eight components of
quantities $j^{l},$ and $S^{l}$, \ $l=0,1,2,3$. Connection between the
4-vector $j^{i}$ and intermediate parameters $A,\eta ^{\alpha }$ has the
form 
\begin{equation}
\begin{array}{c}
j^{0}=A^{2}\cosh \eta \\ 
j^{\alpha }=A^{2}v^{\alpha }\sinh \eta ,\qquad \alpha =1,2,3
\end{array}
\label{a3.13}
\end{equation}
where 
\begin{equation}
{\bf v}=\{v^{\alpha }\},\qquad v^{\alpha }=\eta ^{\alpha }/\eta ,\qquad
\alpha =1,2,3;\qquad {\bf v}^{2}=1.  \label{a3.14}
\end{equation}
The unit 3-pseudovector ${\bf \xi }$ is connected with the spin
4-pseudovector $S^{l}$ by means of the relations 
\begin{equation}
\xi ^{\alpha }=\rho ^{-1}\left[ S^{\alpha }-\frac{j^{\alpha }S^{0}}{%
(j^{0}+\rho )}\right] ,\qquad \alpha =1,2,3;\qquad \rho \equiv \sqrt{%
j^{l}j_{l}}  \label{f1.15}
\end{equation}
\begin{equation}
S^{0}={\bf j\xi },\qquad S^{\alpha }=\rho \xi ^{\alpha }+\frac{({\bf j\xi }%
)j^{\alpha }}{\rho +j^{0}},\qquad \alpha =1,2,3  \label{f1.16}
\end{equation}

Let us make a change of variables in the action (\ref{f1.0}), using
substitution (\ref{f1.11}), (\ref{f1.12}), (\ref{f1.13}). Calculations are
rather bulky, and we omit them (detailed calculation one can find in \cite
{R1995}, or in \cite{R01}). Result of substitution has the form 
\begin{equation}
{\cal S}_{{\rm D}}:\qquad {\cal A}_{D}[j,\varphi ,\kappa ,{\bf \xi }]=\int 
{\cal L}d^{4}x,\qquad {\cal L}={\cal L}_{cl}+{\cal L}_{q1}+{\cal L}_{q2}
\label{c4.15}
\end{equation}
\begin{equation}
{\cal L}_{cl}=-m\rho -\hbar j^{i}\partial _{i}\varphi +\hbar
j^{s}\varepsilon _{iklm}\mu ^{i}\partial _{s}\mu ^{k}z^{l}f^{m},\qquad \rho
\equiv \sqrt{j^{l}j_{l}}  \label{c4.16}
\end{equation}
\begin{equation}
{\cal L}_{q1}=2m\rho \sin ^{2}({\frac{\kappa }{2}})-{\frac{\hbar }{2}}%
S^{l}\partial _{l}\kappa ,  \label{c4.17}
\end{equation}
\begin{equation}
{\cal L}_{q2}=-\hbar \rho \varepsilon _{iklm}q^{i}(\partial ^{k}q^{l})\nu
^{m}  \label{c4.18}
\end{equation}
where the following designations are used 
\begin{equation}
f^{i}=\{1,0,0,0\},\qquad z^{i}=\{0,z^{1},z^{2},z^{3}\}  \label{c4.19}
\end{equation}
\begin{equation}
\nu ^{i}=\xi ^{i}-(\xi ^{s}f_{s})f^{i},\qquad i=0,1,2,3;\qquad \nu ^{i}\nu
_{i}=-1,  \label{c4.20}
\end{equation}
\begin{equation}
\mu ^{i}\equiv \frac{\nu ^{i}}{\sqrt{-(\nu ^{l}+z^{l})(\nu _{l}+z_{l})}}=%
\frac{\nu ^{i}}{\sqrt{2(1-\nu ^{l}z_{l})}}=\frac{\nu ^{i}}{\sqrt{2(1+{\bf %
\xi z})}}.  \label{c4.21}
\end{equation}
\begin{equation}
q^{i}\equiv \frac{j^{i}+f^{i}\rho }{\sqrt{(j^{l}+f^{l}\rho )(j_{l}+f_{l}\rho
)}}=\frac{j^{i}+f^{i}\rho }{\sqrt{2\rho (\rho +j^{l}f_{l})}},\qquad
q_{s}q^{s}=1  \label{c4.22}
\end{equation}
The 4-pseudovector $S^{i}$ is defined by the relation (\ref{f1.16}).

Lagrangian density (\ref{c4.15}) -- (\ref{c4.18}) appears to be
relativistically invariant with respect to quantities $f^{i},z^{i},\xi
^{i},\nu ^{i},j^{i},q^{i},\kappa ,\varphi $, provided the quantity $f^{i}$, $%
i=0,1,2,3$ are considered to be components of a constant timelike unit
4-vector and $z^{i}$ the constant 4-pseudovector orthognal to $f^{i}$. Then
quantities $\xi ^{i}$ and $\nu ^{i}$, defined by (\ref{c4.20}) form
4-pseudovectors. Quantities $\mu ^{i}$ and $q^{i}$ appear to be respectively
4-vector and 4-pseudovector. The quantities $\varphi ,\kappa $ are scalar
and pseudoscalar respectively. The component of $\xi ^{i}$ parallel to
vector $f^{i}$ appears to be arbitrary and unessential. The 4-vector $z^{i}$
appears to be fictitious.

After eliminating $\gamma $-matrices and representing the Lagrangian density
in a relativistically invariant form (\ref{c4.15}) -- (\ref{c4.18}), one
discovers an additional constant timelike unit 4-vector $f^{i}$. There is
only one possible interpretation of this 4-vector. It describes the
space-time split into space and time. It means that the system ${\cal S}_{%
{\rm D}}$, described by the free Dirac equation, is not relativistic, i.e.
it is incompatible with the relativity principles. This result agrees with
the result of the two-dimensional space-time consideration.

\section{Concluding remark}

Dirac equation is a very important equation. It is one of fundamental
equations of quantum electrodynamics. Nonrelativistic character of Dirac
equation means that it is not correct, and one needs a revision.
Fortunately, nonrelativistic description concerns only internal degrees of
freedom connected with the spin variables ${\bf \xi }$ \cite{R01}. At the
low energy processes these degrees of freedom are not excited, and variables 
${\bf \xi }$, describing them, are considered to be constants. At the high
energy processes correction of the Dirac equation may appear to be essential.

\end{document}